# Space Layout of Low-entropy Hydration Shells Guides Protein Binding

Lin Yang[a,b,1,*], Shuai Guo[a,1], Chengyu Hou[c,1], Chencheng Liao[c,1], Jiacheng Li[a], Liping Shi[a], Xiaoliang Ma[a], Shenda Jiang[a], Bing Zheng[d], Yi Fang[e], Lin Ye[b], Xiaodong He[a,f,*]

[a] *National Key Laboratory of Science and Technology on Advanced Composites in Special Environments, Center for Composite Materials and Structures, Harbin Institute of Technology, Harbin 150080, China*

[b] *School of Aerospace, Mechanical and Mechatronic Engineering, The University of Sydney, NSW 2006, Australia*

[c] *School of Electronics and Information Engineering, Harbin Institute of Technology, Harbin 150080, China*

[d] *Key Laboratory of Functional Inorganic Material Chemistry (Ministry of Education) and School of Chemistry and Materials Science, Heilongjiang University, Harbin 150001, P. R. China.*

[e] *Mathematical Science Institute, The Australian National University, Canberra, ACT 0200, Australia.*

[f] *Shenzhen STRONG Advanced Materials Research Institute Co., Ltd, Shenzhen 518035, P. R. China.*

## Abstract

Protein-protein binding enables orderly and lawful biological self-organization, and is therefore considered a miracle of nature. Protein-protein binding is steered by electrostatic forces, hydrogen bonding, van der Waals force, and hydrophobic interactions. Among these physical forces, only the hydrophobic interactions can be considered as long-range intermolecular attractions between proteins in intracellular and extracellular fluid. Low-entropy regions of hydration shells around proteins drive hydrophobic attraction among them that essentially coordinate protein-protein docking in rotational-conformational space of mutual orientations at the guidance stage of the binding. Here, an innovative method was developed for identifying the low-entropy regions of hydration shells of given proteins, and we discovered that the largest low-entropy regions of hydration shells on proteins typically cover the binding sites. According to an analysis of determined protein complex structures, shape matching between the largest low-entropy hydration shell region of a protein and that of its partner at the binding sites is revealed as a regular pattern. Protein-protein binding is thus found to be mainly guided by hydrophobic collapse between the shape-matched low-entropy hydration shells that is verified by bioinformatics analyses of hundreds of structures of protein complexes. A simple algorithm is developed to precisely predict protein binding sites.

## 1. Introduction

*Corresponding author. E-mail address: linyang@hit.edu.cn (Lin Yang) Hexd@hit.edu.cn (Xiaodong He) [1]These authors contributed equally to this work.

Proteins serve a variety of important functions in organisms. A protein's intrinsic biological functions are normally expressed via precise binding with another protein (i.e. a ligand) that derives from the physical phenomenon of protein docking, by which a protein can find its partner protein to form their functional complex structure. Protein-protein binding is a spontaneous physical contact of high specificity established between two specific protein molecules, and erroneous protein-protein binding is highly rare in intracellular and extracellular fluid (*1*). Thus, the physical mechanism responsible for protein-protein binding can be considered the most important mechanism of biological self-organization, functionalization, and diversity.

Protein-protein binding is one of the miracles of nature that human technology finds quite difficult to follow, due to the very large number of possibilities of the rotational-conformational space of mutual orientations potentially sampled by a pair of proteins as they interact. The protein-protein docking is the prediction of the structure of the complex, given the structures of the individual proteins (*2, 3*). Research on protein-protein docking has become more popular due to its potential to predict protein-protein interactions (PPIs) (*4, 5*). In the field of structural biology, protein docking research focuses on computationally simulating the molecular recognition process. A variety of conformational search strategies have been applied to predict protein docking (*6*)(*7*). Although searching algorithms normally aim to achieve an optimized conformation for the complex of a pair of proteins with the minimized free energy of the overall system, sampling of the conformational space in protein-protein docking is still a challenging (*8, 9*).

Protein-protein binding is mainly governed by electrostatic forces, hydrogen bonding, van der Waals forces, and hydrophobic interactions. Among these physical forces, only the hydrophobic interactions can be viewed as long-range intermolecular attractions between proteins in aqueous solutions. The hydration shell (i.e. hydration layer) around a protein has been experimentally found to have dynamics distinct from the bulk water to a distance of 1nm (*10, 11*). Water molecules slow down greatly when they enter the hydration shell of a protein, resulting in lower entropy levels within the shell than bulk water molecules (*10-13*). Protein surface hydrophilic groups are normally hydrogen bonded with surrounding strong polar water molecules in the hydration shell, thereby preventing the surface hydrogen bond donors of a protein from randomly hydrogen-bonding or electrostatic attracting with the hydrogen bond acceptors of another protein, namely, preventing erroneous protein-protein binding in unsaturated aqueous solution (*11, 14-18*). Thus, protein-protein binding phenomenon should start from the long-range hydrophobic attraction between low-entropy regions within the protein's hydration shell (*19, 20*). Despite this, there have been few studies to accurately identify low-entropy regions of protein's hydration shell.

## 2 Low-entropy regions of hydration Shells of proteins

Recent experimental evidence has shown that the protein surface hydration dynamics are highly heterogeneous over the global protein surface (*21-23*). This indicate the existence of low-entropy regions of protein hydration shells. The distribution of hydrophilic and hydrophobic groups on a protein surface determines the protein surface hydration dynamics (*21, 22*). In experimentally determined protein structures, there are always some exposed hydrophilic backbone carbonyl oxygen atoms and backbone amide hydrogen atoms at the protein surfaces. While some surface hydrophobic side-chains protrude outward to surrounding water molecules that may shield the hydrophilic backbone carbonyl oxygen atoms and backbone amide hydrogen atoms. For example, according to all hydrophobicity scales, isoleucine, valine, and leucine residues are highly hydrophobic, even if the residues contain hydrophilic backbone carbonyl oxygen atoms and amide hydrogen atoms (see Fig.1a). Hydrophobic side-chains of isoleucine, valine, and leucine residues on protein surfaces expel surrounding water molecules to van der Waals interactions operating distances (0.3 to 0.6 nm) to form the low-entropy hydration shells (i.e. ordered water molecules cages). van der Waals interactions operating distances are much larger than the hydrogen bonding distance (0.3nm). In this way, the ordered water molecules in the low-entropy hydration shells are inhibited from fluctuating and rearrangement, and are therefore prevented from frequently hydrogen bonding with the backbone carbonyl oxygen atoms and amide hydrogen atoms, as shown in Fig.1a. This indicates that the highly hydrophobic side-chains can shield the hydrophilicity of the backbone atoms to a certain extent. This make hydration shells covering the backbone carbonyl oxygen atoms and amide hydrogen atoms of these hydrophobic residues can be regarded as low-entropy hydration shells, due to the hydrogen-bond rearrangements between the backbone hydrophilic atoms and surrounding water molecules are inhibited (see Fig1.a). It is worth noting that the side-chains of tyrosine, tryptophan, cysteine, methionine, phenylalanine, lysine, arginine and alanine residues also contain highly hydrophobic structures (i.e. alkyl and benzene ring). The hydration shells surrounding the backbone carbonyl oxygen groups and the amide hydrogen groups of these residues also should be considered low-entropy hydration shells (see Fig1).

According to different hydrophobicity scales of amino acid residues, tryptophan and tyrosine exhibit different hydrophilic and hydrophobic properties (*24-27*). It has previously been noted that tryptophan and tyrosine amino acid only express their hydrophilicity via a tiny CO or NH group in their long side-chains, whereas the other portions of the side-chains are highly hydrophobic alkyl and benzene ring structures (see Fig.1b) (*28*). The characteristic of hydration shell water molecules surrounding hydrophobic groups is that their hydrogen bonding network is much more ordered than free liquid water molecules, that is, their entropy is lower (less entropy in the system) (*29-33*). Therefore, ordered water molecules are fixed in low-entropy hydration shells around the highly hydrophobic alkyl and benzene ring structures and are expelled to van der Waals interactions operating distances. The tiny CO or NH group is most likely to hydrogen bond with the hydrophobic-group-induced low entropy water cages rather than destroy the ordered water molecules network (see Fig.1b). For instance, the hydrogen-bond rearrangements between water molecules and the hydrophilic NH group of tyrosine side-chain can be inhibited by the hydrophobic benzene ring of the side-chain, due to the NH-group's neighboring water molecules were already fixed in the ordered network. This explains why tryptophan and tyrosine are categorized as hydrophobic residues in some hydrophobicity scales (*25-27*). Therefore, hydration shells surrounding the side-chains of tryptophan, tyrosine, lysine can be regarded as a low entropy hydration shell (see Fig.1b). When the side-chains of tryptophan, tyrosine, lysine locate at the surface of a protein, their hydration shells should be considered low-entropy as whole.

It is important to note that proteins normally have an abundance of intramolecular hydrogen bonds. For example, protein secondary structures arise from the hydrogen bonds formed between the amide proton and the carbonyl oxygen of the polypeptide backbone. To form these intramolecular hydrogen bonds, nascent unfolded polypeptide chains need to escape from hydrogen bonding with surrounding polar water molecules that require entropy-enthalpy compensations during the protein folding, according to the Gibbs free energy equation. The entropy-enthalpy compensations are initially driven by laterally hydrophobic collapse among the side-chains of adjacent residues in the sequences of unfolded protein chains (*28*). As a result of the entropy-enthalpy compensations, water molecules cannot break the intramolecular hydrogen bonds of proteins by competing with the donors and acceptors of these intramolecular hydrogen bonds (*28*). It means that protein intramolecular hydrogen bonds saturate many of the hydrogen bonds formed by surface hydrophilic groups of proteins. Therefore, we can consider that these intramolecular hydrogen-bonded hydrophilic groups can't destroy surrounding hydrophobic-group-induced low-entropy water molecules network, due to their hydrophilicity have been expressed by the intramolecular hydrogen bonds. On protein surfaces, the hydration shells covering these intramolecular hydrogen-bonded hydrophilic groups can be considered as low-entropy hydration shells, due to hydrophobic alkyl always neighbored with the hydrophilic groups in protein structures (see supplementary Fig.S1). In secondary structures, for instance, the hydration shells covering the hydrogen bonded backbone amide proton and carbonyl oxygen should be regarded as low-entropy structures (see Fig.2). Based on this finding, we find out that typical secondary structures normally characterized by their one-side surfaces fully covered by low-entropy hydration shells. This enable secondary structures fully hydrophobic collapse together to form the protein tertiary structure, as illustrated in Fig.2.

Based on the above analysis, we can judge the low-entropy nature of the hydration shells surrounding specific surface hydrophilic groups of proteins. Consequently, we are able to map low-entropy regions of the hydration shell of a given protein (see Fig.3). We map the low-entropy hydration shell regions for 100 protein pairs, and found out that binding sites on proteins are typically covered by the largest low-entropy regions of the proteins hydration shells. All the 100 protein complexes were randomly selected from the PDB. The shape of the largest low-entropy hydration shell region of a given protein can be easily achieved from their projective images (*28*). Surprisingly, we found out that shape matching between the largest low-entropy hydration shell region of a protein and that of its partner at the binding sites is prevailing in all the tested protein complexes. For example, the shape and size matching of the largest low-entropy hydration shell regions at the binding sites of the spike protein of the omicron variant of the severe acute respiratory syndrome coronavirus (SARS-CoV-2) and the receptor angiotensin converting enzyme 2 (ACE2) is demonstrated in Fig.3.

The shapes and sizes of the largest low-entropy regions of protein hydration shells can be used as parameters to predict protein docking. Through analyzing the hydrophobic attraction relationships among low-entropy regions of protein hydration shells of hundreds of protein pairs, we find out that the binding sites of a pair of proteins are always characterized by two rules of the space layout of low-entropy regions of the hydration shells at the binding sites. First, the docking position maximizes the overlapping of the largest low-entropy hydration shell region of a protein and that of its partner. Secondly, the binding sites of a pair of proteins must allow sufficient interfacial contact at the docking position of the complex. Ordered water molecules fixed in the water cages of the low-entropy regions of hydration shells that drive hydrophobic collapse of the low-entropy hydration shell regions in-between proteins, thereby rearrange ordered water molecules to free liquid water molecules to increase entropy. The bind affinity between two proteins are initially resourced from long-range hydrophobic effect among the low-entropy hydration shell regions of the two proteins at the binding sites, enable the shape-matched largest low-entropy hydration shell regions fully collapse in-between the two proteins (*34-37*). To prove that protein-protein binging process is guided by hydrophobic attraction between shape-matched largest low-entropy hydration shells of the proteins, we try to predict the binding sites of the 100 protein complexes by using the above two rules (see Figure 4 and Supplementary). All the binding sites of the 100 protein complexes were successfully predicted by the using the two rules, which provides potent proof for the theory of hydrophobic collapse between the shape-matched largest low-entropy regions of the hydration shells. All the 100 protein complex were randomly selected from the PDB.

With further hydrogen bonding matches, protein-protein docking can be accurately predicted.

## 5. Conclusion

As a typical spontaneous reaction, the guidance stage of protein-protein binding must release Gibbs free energy as it proceed. The early steps of protein-protein binding should be not dominated by electrostatic interaction or hydrogen bonding in-between the two proteins, due to the shielding effect of polar water molecules of the hydration shells. A protein-protein binding begins with a long-range hydrophobic attraction between the low-entropy regions of hydration shells of individual proteins. Entropy increase caused by the hydrophobic attraction guides the docking process and provides the binding affinity (*28*). By the analysis, we show that all the binding sites of protein pairs were covered by the shape-matched largest low-entropy regions of the hydration shells, enable the largest low-entropy regions fully collapse at the protein-protein interfaces during the docking processes. Protein-protein binding is mainly guided by hydrophobic collapse between shape-matched low-entropy regions of hydration shells of individual proteins. Despite the difficulty in identifying low-entropy hydration shells around a protein using experiment methods, the theoretical approach allows us to identify the largest low-entropy hydration shell areas around proteins that can be used to predict protein binding sites. Space layout of the low-entropy region of the protein's hydration shell acts like a 'Lock or Key' for guiding the protein-protein binding in a precise manner.

**Materials and Methods**

**Protein structures**

In this study, many experimentally determined native structures of proteins are used to study the protein-protein docking mechanism. All the three-dimensional (3D) structure data of protein molecules are resourced from the PDB database. IDs of these proteins according to PDB database are marked in all the figures. In order to show the distribution of low-entropy hydration shells on the surface of proteins at the binding sites in these figures, we used the structural biology visualization software PyMOL to display the low-entropy hydration shell areas.

## Hydrophilicity of residues

The detailed space layout of hydrophilicity of residues can be easily identified from the amount of charge of atoms according to the charmm36 force field (see supplementary) (*38*).

## Acknowledgements

Lin Yang is indebted to Daniel Wagner from the Weizmann Institute of Science and Liyong Tong from the University of Sydney for their guidance. Lin Yang is grateful for his research experience in the Weizmann Institute of Science for inspiration. The authors acknowledge the financial support from the National Natural Science Foundation of China (Grant 21601054), Shenzhen Science and Technology Program (Grant No. KQTD2016112814303055), Natural Science Foundation of Heilongjiang Province (Grant LH2019F017), Science Foundation of the National Key Laboratory of Science and Technology on Advanced Composites in Special Environments, the Fundamental Research Funds for the Central Universities of China and the University Nursing Program for Young Scholars with Creative Talents in Heilongjiang Province of China (Grants UNPYSCT-2017126).

**Author Contributions** L.Yang, L.Ye, and X.H. formulated the study. L.Yang, J. L, S.G., X.M., C.H., H. Z., L.S., B.Z., and C.L. collected and analyzed the PDB data. C.H., L.Yang, C.L. wrote programs. L.Yang, L.Ye, Y.F. and X.H. wrote the paper, and all authors contributed to revising it. All authors discussed the results and theoretical interpretations.

## Additional Information

The authors declare no competing financial interests.

## References

1. X. Du *et al.*, Insights into Protein-Ligand Interactions: Mechanisms, Models, and Methods. *Int J Mol Sci* **17**, 144 (2016).
2. Ilya A. Vakser, Protein-Protein Docking: From Interaction to Interactome. *Biophysical Journal* **107**, 1785-1793 (2014).
3. So Much More to Know &#133. *Science* **309**, 78 (2005).
4. S. Das, S. Chakrabarti, Classification and prediction of protein–protein interaction interface using machine learning algorithm. *Scientific Reports* **11**, 1761 (2021).
5. H. Lu *et al.*, Recent advances in the development of protein–protein interactions modulators: mechanisms and clinical trials. *Signal Transduction and Targeted Therapy* **5**, 213 (2020).

6. G. Jones, P. Willett, R. C. Glen, A. R. Leach, R. Taylor, Development and validation of a genetic algorithm for flexible docking11Edited by F. E. Cohen. *Journal of Molecular Biology* **267**, 727-748 (1997).
7. D. S. Goodsell, G. M. Morris, A. J. Olson, Automated docking of flexible ligands: Applications of autodock. *Journal of Molecular Recognition* **9**, 1-5 (1996).
8. C. J. Camacho, S. Vajda, Protein docking along smooth association pathways. *Proceedings of the National Academy of Sciences of the United States of America* **98**, 10636-10641 (2001).
9. D. Kozakov *et al.*, The ClusPro web server for protein-protein docking. *Nat Protoc* **12**, 255-278 (2017).
10. C. M. Soares, V. H. Teixeira, A. M. Baptista, Protein structure and dynamics in nonaqueous solvents: insights from molecular dynamics simulation studies. *Biophys J* **84**, 1628-1641 (2003).
11. L. Zhang *et al.*, Mapping hydration dynamics around a protein surface. *Proceedings of the National Academy of Sciences of the United States of America* **104**, 18461-18466 (2007).
12. D. Zhong, S. K. Pal, A. H. Zewail, Biological water: A critique. *Chemical Physics Letters* **503**, 1-11 (2011).
13. A. Debnath, B. Mukherjee, K. G. Ayappa, P. K. Maiti, S.-T. Lin, Entropy and dynamics of water in hydration layers of a bilayer. *J. Chem. Phys.* **133**, 174704 (2010).
14. Y. Lin *et al.*, Universal Initial Thermodynamic Metastable state of Unfolded Proteins. *Progress in biochemistry and biophysics* **46**, 8 (2019).
15. B. Qiao, F. Jiménez-Ángeles, T. D. Nguyen, M. Olvera de la Cruz, Water follows polar and nonpolar protein surface domains. *Proceedings of the National Academy of Sciences* **116**, 19274-19281 (2019).
16. A. McPherson, J. A. Gavira, Introduction to protein crystallization. *Acta Crystallogr F Struct Biol Commun* **70**, 2-20 (2014).
17. J. J. Urban, B. G. Tillman, W. A. Cronin, Fluoroolefins as Peptide Mimetics:  A Computational Study of Structure, Charge Distribution, Hydration, and Hydrogen Bonding. *The Journal of Physical Chemistry A* **110**, 11120-11129 (2006).
18. D. Chen *et al.*, Regulation of protein-ligand binding affinity by hydrogen bond pairing. *Science Advances* **2**, e1501240 (2016).
19. J. Li *et al.*, A Hydrophobic-Interaction-Based Mechanism Triggers Docking between the SARS-CoV-2 Spike and Angiotensin-Converting Enzyme 2. **n/a**, 2000067.
20. L. Yang *et al.*, SARS-CoV-2 Variants, RBD Mutations, Binding Affinity, and Antibody Escape. **22**, 12114 (2021).
21. Y. Qin, L. Wang, D. Zhong, Dynamics and mechanism of ultrafast water–protein interactions. **113**, 8424-8429 (2016).
22. D. Laage, T. Elsaesser, J. T. Hynes, Water Dynamics in the Hydration Shells of Biomolecules. *Chemical Reviews* **117**, 10694-10725 (2017).
23. R. Barnes *et al.*, Spatially Heterogeneous Surface Water Diffusivity around Structured Protein Surfaces at Equilibrium. *Journal of the American Chemical Society* **139**, 17890-17901 (2017).
24. J. Kyte, R. F. Doolittle, A simple method for displaying the hydropathic character of a protein. *Journal of Molecular Biology* **157**, 105-132 (1982).
25. Chi-Hao *et al.*, Hydrophobicity of amino acid residues: Differential scanning calorimetry and synthesis of the aromatic analogues of the polypentapeptide of elastin.   (1992).
26. D. J. J. o. N. Dougherty, Cation-π Interactions Involving Aromatic Amino Acids. **137**, 1504S (2007).
27. IMGT standardized criteria for statistical analysis of immunoglobulin V-REGION amino acid properties. %J *Journal of Molecular Recognition*. **17**, 17–32 (2004).
28. J. Li *et al.*, Entropy-Enthalpy Compensations Fold Proteins in Precise Ways. *Int. J. Mol. Sci.* **22**, 9653 (2021).
29. D. Cui, S. Ou, S. Patel, Protein-spanning water networks and implications for prediction of protein–protein interactions mediated through hydrophobic effects. *Proteins: Structure, Function, and Bioinformatics* **82**, 3312-3326 (2014).
30. Q. Wang, C. Smith, Molecular biology genes to proteins, 3rd edition by B. E. Tropp. **36**, 318-319 (2008).
31. J. Grdadolnik, F. Merzel, F. Avbelj, Origin of hydrophobicity and enhanced water hydrogen bond strength near purely hydrophobic solutes. *Proceedings of the National Academy of Sciences* **114**, 322-327 (2017).
32. Q. Sun, The physical origin of hydrophobic effects. *Chemical Physics Letters* **672**, 21-25 (2017).
33. P. Ball, Water is an active matrix of life for cell and molecular biology. *Proceedings of the National Academy of Sciences* **114**, 13327 (2017).
34. J. Li *et al.*, A Hydrophobic-Interaction-Based Mechanism Triggers Docking between the SARS-CoV-2 Spike and Angiotensin-Converting Enzyme 2. *Global Challenges* **4**, 2000067 (2020).
35. A. Berchanski, B. Shapira, M. Eisenstein, Hydrophobic complementarity in protein-protein docking. *Proteins* **56**, 130-142 (2004).
36. E. E. Meyer, K. J. Rosenberg, J. Israelachvili, Recent progress in understanding hydrophobic interactions. **103**, 15739-15746 (2006).
37. C. Chothia, J. Janin, Principles of protein-protein recognition. *Nature* **256**, 705-708 (1975).
38. B. R. Brooks *et al.*, CHARMM: the biomolecular simulation program. *J Comput Chem* **30**, 1545-1614 (2009).

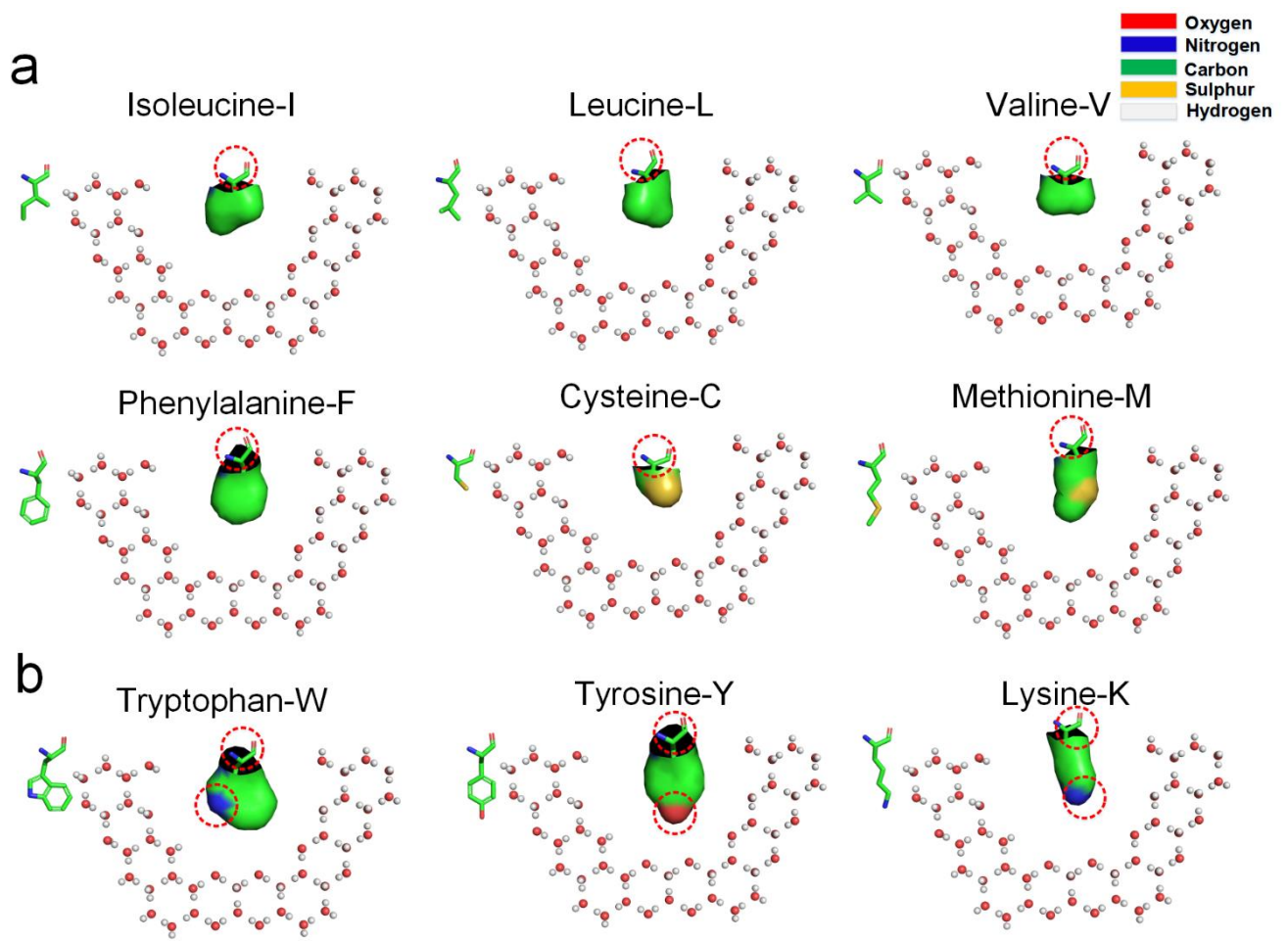

Fig.1 Low-entropy hydration shells of amino acid side-chains (hydrophobic portions are highlighted green and yellow). Hydrophobic groups covered by low-entropy hydration shells are highlighted by red dash circles.

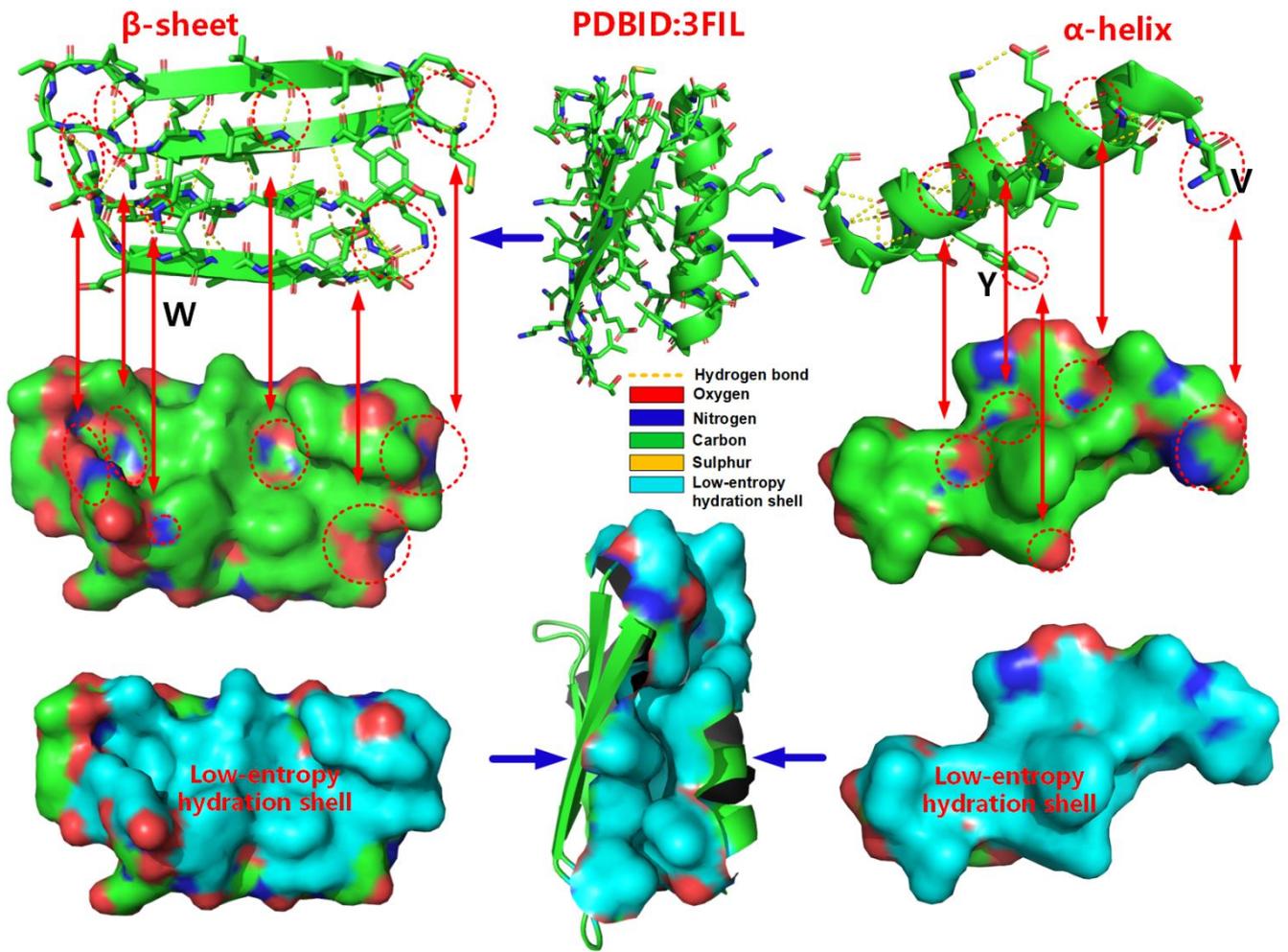

Fig.2 Low-entropy hydration shells cover one-side surfaces of the secondary structures before the folding of the tertiary structure of a protein (PDBID: 3FIL). The low-entropy hydration shell region is highlighted in cyan color.

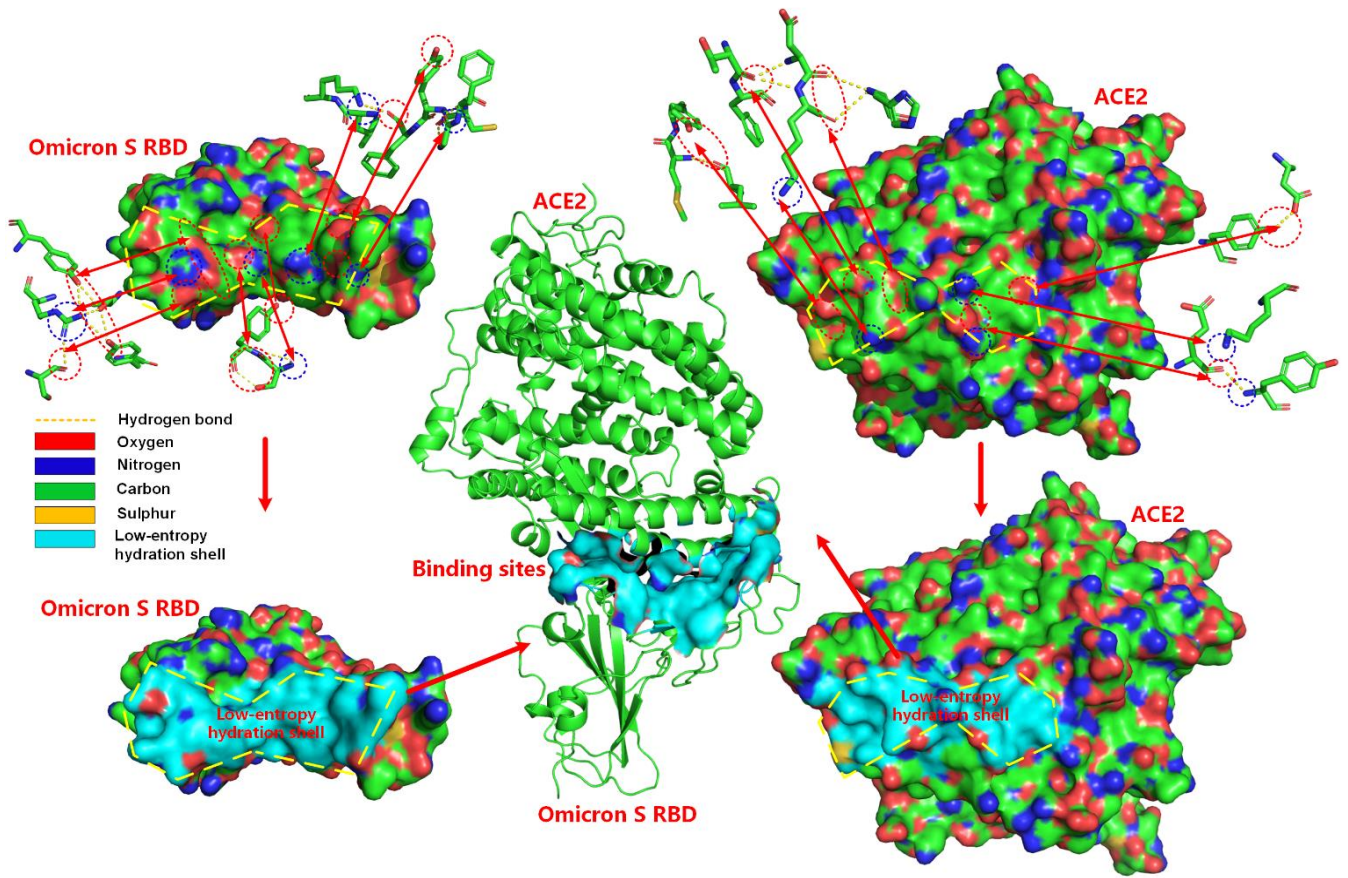

Fig. 3 Low-entropy hydration shells on the binding sites of the RBD of Omicron variant S protein variant and the ACE2. The binding sites of the two proteins are highlight by yellow dash lines, the low-entropy hydration shell region is highlighted in cyan color.

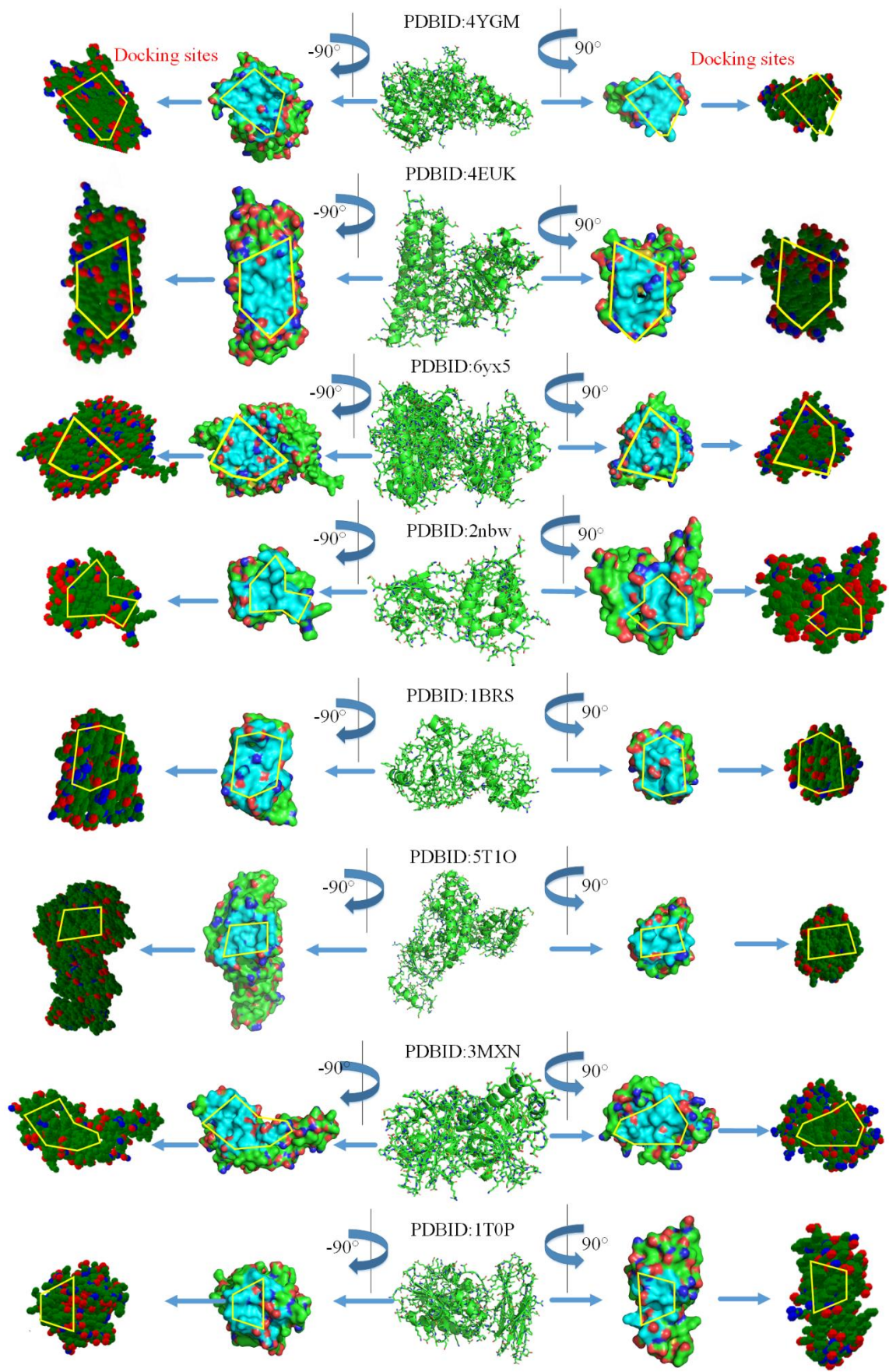

Fig.4 The prediction of binding sites of protein pairs through identifying the largest low-entropy hydration shells of individual proteins.